\documentstyle[a4]{article}

\input amssym.def
\input amssym

\newcommand{\seZ}{\Bbb Z}
\newcommand{\wek}[2]{\newcommand{#1}{\mbox{\bf #2}}}
\wek{\vA}{A} \wek{\vR}{R} \wek{\vH}{H} 
\wek{\vp}{p} \wek{\vr}{r} \wek{\vh}{h}
\wek{\va}{a} \wek{\vq}{q} \wek{\vk}{k}
\newcommand{\ii}{\mbox{\rm i}}
\newcommand{\de}[1]{\delta_{#1}} \newcommand{\req}[1]{(\ref{#1})}
\newcommand{\ie}{{\it i.e.}}     \newcommand{\eg}{{\it e.g.}}
\newcommand{\half}{{\textstyle{1\over2}}}

\begin{document}

\title{Kronecker products of projective representations
of translation groups\thanks{This work was presented as a poster
at the Fifth International Wigner Symposium, Vienna, August 25--29
1997; the author's participation in the symposium was partially sponsored by
the Stefan Batory Foundation.}}
\author{Wojciech FLOREK\\
\date{Adam Mickiewicz University, Institute of Physics,
Computational Physics Div.,
Umultowska 85, 61--614 Pozna\'n, Poland\\}
(florek@phys.amu.edu.pl)}
\date{23 September 1997}

\maketitle

\begin{abstract}
 Irreducible projective representations of the translation group of a
finite $N\times N$ two-dimensional lattice can be labeled by symbols
$\langle n,l;\mbox{\bf q}\rangle$, where $N=\nu n$, $\gcd(l,n)=1$ and
{\bf q} denotes an irreducible representation of $Z_\nu^2$. Obtained
matrices are $n$-dimensional and the factor system of this
representation does not depend on {\bf q} and equals 
$m_n^{(l)}([n_1,n_2],[n'_1,n'_2])=\exp(2\pi{\rm i}\, l n_2n'_1/N)$.
For a given $n$ the $N\times N$ lattice can be viewed as a
$\nu\times\nu$ lattice consisting of $n\times n$ magnetic cells.  The
Kronecker product of such representations is another projective
representation which can be decomposed into irreducible ones. It is
interesting that such product can lead to the magnetic periodicity
different from $n$, $n'$ and even $nn'$ or ${\rm lcm}(n,n')$. For
example, a product $\langle n,l;\mbox{\bf q}\rangle\otimes\langle
n,l;\mbox{\bf q}'\rangle$ for {\em even} $N$ decomposes into
representations $\langle {n\over2},l;\mbox{\bf k}\rangle$:  there are
four representations with $k_i-q_i-q'_i=0,\nu$ and each of them
appears $n/2$ times. Similarly, the coupling of $d$ representations
$\langle n,1;\mbox{\bf q}^{(j)}\rangle$, $j=1,2,\dots,d$ with $n=dM$
changes the magnetic period from $n$ to $M$. It is caused by
multiplication of the {\em charge} by $d$ what corresponds to a
system of $d$ electrons.
 \end{abstract}
\section{Introduction} 

Projective (or ray) representations investigated by Schur at the beginning
of this century \cite{schur} are widely applied in quantum mechanics (due
to factor systems related with them) and crystallography (especially as a
tool in construction of space group representations). In the sixties Brown
\cite{brown} applied them to investigation of movement of a Bloch electron
in a magnetic field. Almost at the same time Zak \cite{zak1} proposed an
equivalent approach: ray representations of the translation group were
considered as vector (\ie\ ordinary) representations of its covering group
being in fact a central extension of the translation group and a group of
factors (see also \cite{florek94}). The problem investigated by these
authors is strongly related to the Landau quantization and, therefore, to
the quantum Hall effect (see, for example, articles by Zak \cite{zak2},
Dana, Avron and Zak \cite{DAZ}, Aoki \cite{aoki}). Many authors, however,
rejected some representations, obtained in the mathematical analysis of this
problem, and claimed that they are `nonphysical' \cite{zak1}. 

In the series of articles \cite{florek94,florek96} the magnetic translation
group was studied within the frame proposed by Zak, \ie\ as a central
extension of the translation group. These investigations gave a basis to
construct and consider all representations, including those called
`nonphysical' \cite{florek97a}. Moreover, the physical relevance and
possible applications of such representations were indicated.

In this article the magnetic translation operators are studied applying
Brown's approach, \ie\ projective representations of the (two-dimensional)
translation group are considered. In Sec.~2, after recalling Brown's
definitions, matrix elements of finite irreducible projective representations
are given (they have been obtained from those introduced by Brown applying a
gauge transformation). In the next section tensor (Kronecker) products of
such representations are investigated.

\section{Irreducible projective representations}

Brown \cite{brown} introduced magnetic translation operators as
\begin{equation}\label{mto}
 T(\vR)=\exp[-\ii\vR\cdot(\vp-e\vA/c)/\hbar]\,.
 \end{equation}
 These operators commute with the Hamiltonian
 \begin{equation}
 {\cal H}={1\over{2m}}(\vp+e\vA/c)^2+V(\vr)\,,
 \end{equation}
 describing an electron in a periodic potential $V(\vr)$ and a uniform
magnetic field
\begin{equation}
 \vH={\bf rot}\,\vA\,,\qquad\mbox{where}\qquad \vA=\half(\vH\times\vr)\,.
 \end{equation}
 The introduced operators form a projective representation of the crystal
translation groups, what is expressed by the following relation:
\begin{equation}
 T(\vR)T(\vR')=T(\vR+\vR')m'(\vR,\vR')\,,
 \end{equation}
 where
 \begin{equation}
 \label{fac}
 m'(\vR,\vR')=\exp[-\ii(\vR\times\vR')\cdot\vh/2]
 \end{equation}
 is a factor system of this representation with $\vh=e\vH/\hbar c$. 
 
Imposing the periodic conditions Brown showed (see also \cite{zak1}) that
the magnetic field can be assumed to equal
\begin{equation}\label{cond}
 \vh =\frac{2\pi}{N} \frac{L}{\Omega} \va_3
 \end{equation}
 for an integer $L$ mutually prime with $N$; $\Omega
=(\va_1\times\va_2)\cdot\va_3$ is the volume of the primitive cell, $\va_i$
are the primitive translations and $N$ the period of the crystal lattice.
For such a choice of $\vh$ [and factor system \req{fac}] Brown obtained
$N$-dimensional irreducible projective representations for
$\vR=n_1\va_1+n_2\va_2$ with matrix elements
\begin{equation}\label{brep}
 D_{jk}(\vR)=\exp\left[\frac{\pi\ii}{N} Ln_1(n_2+2j)\right]\de{j,k-n_2}\,;
 \pmod{N}\,;\qquad j,k=0,1,\dots,N-1\,.
 \end{equation}
 The factor system for this representation agrees with that given by
\req{fac} since
 $$
 D(\vR)D(\vR')=D(\vR+\vR')
 \exp\left[\frac{\pi\ii}{N} L(n_2n'_1-n_1n'_2)\right]\,.
 $$
 Note that all factors are roots of 1 of the order $2N$, whereas the
dimension of the considered representations is $N$. Therefore, there exists
an equivalent normalized (and standard) factor system $m$, \ie\ such a
system that all factors are the $N$-th roots of 1 \cite{alt}. It can be
obtained if each matrix $D(\vR)$ will be multiplied by $\phi(\vR)
=\exp(-\pi\ii Ln_1n_2/N)$. In this way new (and nonequivalent to the
previous ones) irreducible representations are obtained
\begin{equation}\label{DNL}
 \langle N,L;{\bf0}\rangle_{jk}[n_1,n_2]
 =\de{j,k-n_2}\omega_N^{Ln_1j}\,,\qquad j,k=0,1,\dots,N-1\,,\quad
 \omega_N=\exp(2\pi\ii/N)\,,
 \end{equation}
 where $\langle N,L;{\bf0}\rangle$ denotes the representation (the role of
the zero vector {\bf0} will be explain below) and a vector
$n_1\va_1+n_2\va_2$ was replaced by a pair $[n_1,n_2]$. Since only the
unique element $[0,0]$ has nonzero character, then this representation is
irreducible (as projective representation of the translation group
$T_N\simeq\seZ_N\otimes\seZ_N$). It is easily shown that a factor system
\begin{equation}\label{fNL}
 m_N^{(L)}([n_1,n_2],[n'_1,n'_2])=\omega_N^{L n_2n'_1}
 \end{equation}
 corresponds to this form of representations. It can be shown (see
\cite{florek97b}; more detailed discussion of different gauges is in
preparation) that this factor system corresponds to the Landau gauge, used
in many papers (see, \eg, \cite{aoki,WZ}). Three important facts should be
stressed:
\begin{itemize}
 \item Projective representations with different, though equivalent, factor
systems are {\it nonequivalent} \cite{alt}, so representations discussed in
this paper and those introduced by Brown are nonequivalent. However, the
same set of basis function can be used.
 \item In fact, modification of the factor system \req{fac} corresponds to
different choice of the gauge $\vA$; it was shown \cite{florek97b} how to
introduce the magnetic translations for any gauge $\vA$ (such that $\vH={\bf
rot}\,\vA$).
 \item Equivalent factor systems lead to the same expression for the commutator
  $$
 D(\vR)D(\vR')D^{-1}(\vR)D^{-1}(\vR')=\omega_N^{-L (n_1n'_2-n_2n'_1)}\,.
  $$
\end{itemize}

 The actual form of basis function is not discussed here (see, \eg,
\cite{brown,zak1,zak2,dana1} for more details). It is worth noting that
these functions, denoted as $|s\rangle$ with $s=0,1,\dots,N-1$, are
eigenfunctions of $\langle N,L;{\bf0}\rangle[n_1,0]$ operators, whereas the
operators $\langle N,L;{\bf0}\rangle[0,n_2]$ permutes them in a cyclic way
({\it cf.} \cite{florek94,sch}):
 \begin{eqnarray}\label{s1}
 \langle N,L;{\bf0}\rangle[n_1,0]|s\rangle 
 &=& \omega_N^{Ln_1s}|s\rangle\,;\\
 \langle N,L;{\bf0}\rangle[0,n_2]|s\rangle &=& |s-n_2\rangle\,\pmod{N}.
 \label{s2}
 \end{eqnarray}
 The special choice of $\langle N,L;{\bf0}\rangle$ put $\va_1$ and $\va_2$
on a different footing.

Let us assume that the number $L$ has a common factor with $N$, say $L=l\nu$
and $N=n\nu$ with $\nu=\gcd(L,N)>1$ (please recall that $L$ is simply
connected with the magnetic field magnitude). It is easy to notice that the
{\it magnetic} periodicity is obtained for the smaller period $n$ and the
factor system \req{fNL} can be written as
\begin{equation}\label{fnl}
 m_n^{(l)}([n_1,n_2],[n'_1,n'_2])=\omega_n^{l n_2n'_1}\,.
 \end{equation}
 Therefore, one may consider factor systems \req{fnl} for all divisors $n$
of $N$ and $l$ mutually prime with $n$ (\ie\ $\gcd(n,N)=n$ and
$\gcd(l,n)=1$). It is an easy task of combinatorics to show that $N$
different factor systems, corresponding to $L=0,1,\dots,N-1$, are obtained
in this way \cite{kerb}. Since even for $n<N$ the $N\times N$ lattice is
still under the question, so $N$ will be called hereafter the {\it
crystal}\/ period, whereas $n$, for which $T(n\vR)={\bf1}$, will be called
the {\it magnetic}\/ period. Hence, the $N\times N$ lattice can be viewed as
a $\nu\times\nu$ lattice, with the translation group
$T_\nu=\seZ_\nu\otimes\seZ_\nu$, of $n\times n$ magnetic cells. Let
$\vq=[q_1,q_2]$ and 
\begin{equation}\label{q}
 \langle\vq\rangle_\nu[\xi_1,\xi_2]
 =\exp[-2\pi\ii(q_1\xi_1+q_2\xi_2)/\nu]
 =\omega_{\nu}^{-(q_1\xi_1+q_2\xi_2)}
 \end{equation}
 be the irreducible representation of $T_\nu$. Then it is easy to check that
$n$-dimensional matrices
\begin{equation}\label{Dnlq}
 \langle n,l;\vq\rangle_{jk}[n_1,n_2]
 =\de{j,k-\eta_2}\omega_n^{l\eta_1j}\omega_{\nu}^{-(q_1\xi_1+q_2\xi_2)}
 \end{equation}
 form a projective irreducible representation of the group $T_N$ with a
factor system \req{fnl}. In this formula $[\xi_1,\xi_2]$ labels magnetic
cells, whereas $[\eta_1,\eta_2]$ labels positions within a magnetic cell,
\ie\ $n_i=\eta_i+\xi_in$. As the basis function the eigenvectors
$|n,l;\vq;s\rangle$, $0\le s<n$, of the matrix $\langle n,l;\vq\rangle[1,0]$
will be used. The character of the representation \req{Dnlq} is easily
calculated as 
 \begin{equation}\label{chinlq}
 \chi\langle n,l;\vq\rangle[n_1,n_2]
 =\de{\eta_1,0}\de{\eta_2,0}n \omega_{\nu}^{-(q_1\xi_1+q_2\xi_2)}\,.
 \end{equation}
 For given $n$ and $l$ (\ie\ for a given factor system) we obtained $\nu^2$
nonequivalent irreducible projective representations (labeled by $\vq$), so
we obtained all of them \cite{alt}. In particular we have
 $$
 \langle1,1;\vq\rangle=\langle\vq\rangle_N,.
 $$

\section{Products of irreducible projective representations}

\noindent Let us consider a product of two projective representations $T$ 
and $T'$ of
a given group $G$ with factors systems $m$ and $m'$. For a
matrix element of the considered product we have $(T\otimes T')_{ij,kl}(g)=
T_{ik}(g)T'_{jl}(g)$ so
 \begin{eqnarray}
 [(T\otimes T')(g)(T\otimes T')(g')]_{ij,kl}
 &=&\sum_{p,q}(T\otimes T')_{ij,pq}(g)(T\otimes
T')_{pq,kl}(g')\nonumber\\
 &=&\sum_{p,q} T_{ip}(g)T'_{jq}(g) T_{pk}(g')T'_{ql}(g') \nonumber\\
 &=& m(g,g')m'(g,g') T_{ik}(gg')T'_{jl}(gg') \nonumber\\
 &=& m''(g,g') (T\otimes T')_{ij,kl}(gg')\,,
 \end{eqnarray}
 where $m''(g,g')=m(g,g')m'(g,g')$ is, in general, a new factor system. (Of
course a product of two vector representations is a vector representation.)
In the considered case all factors are the $N$-th root of 1 and a product of
two factors $m_n^{(l)}$ and $m_{n'}^{(l')}$, is equal to 
\begin{equation}\label{fprod}
 m([n_1,n_2],[n'_1,n'_2])=\omega_N^{(l\nu+l'\nu')n_2n'_1}\,,
 \end{equation}
 so it corresponds to representation with $L=l\nu+l'\nu'$. It means that the
set of factor systems \req{fnl} is closed with respect to the multiplication
and, therefore, the representations \req{Dnlq} and their direct sums form a
closed set with respect to the tensor product. Of course, we can add
representations with the same factor system only and vice versa --- if
a given projective representation with factor system $m$ is reducible then
it can be decomposed into a direct sum of irreducible projective
representations with the same factor system $m$. Moreover, the orthogonality
relations for representations and their characters are valid for
representations with the same factor system~\cite{alt}. Hence, one must be
very careful decomposing a given projective representation ---
representations with different factor systems {\it can not} be compared.
 
Let $D$ be a product of two irreducible representations $\langle
n,l;\vq\rangle$ and $\langle n',l';\vq'\rangle$. Its factor system is given
by \req{fprod}, its dimension equals $nn'$ and its character is
\begin{equation}
 \chi^D[n_1,n_2]
 =\de{\eta_1,0}\de{\eta_2,0}\de{\eta'_1,0}\de{\eta'_2,0}nn'
 \omega_N^{-n(q_1\xi_1+q_2\xi_2)-n'(q'_1\xi'_1+q'_2\xi'_2)}\,,
 \end{equation}
 so it is nonzero only for $n_i=x_i m$, where $m=nn'/\gamma$,
$\gamma=\gcd(n,n')$, $0\le x_i<\mu=N/m=\gcd(\nu,\nu')$. Substituting $m$ and
$\mu$ to the above formula one easily obtains
\begin{equation}\label{chiD}
 \chi^D[n_1,n_2]
 =\de{\eta_1,0}\de{\eta_2,0} m\gamma
 \omega_{\mu}^{-(q_1+q'_1)x_1-(q_2+q'_2)x_2}\,;\pmod m.
 \end{equation}
 Since $\nu/\mu=n'/\gamma$ then $L$ in \req{fprod} can be written as
 \begin{equation}\label{fprm}
 L=\mu\left(\frac{l\nu}{\mu}+\frac{l'\nu'}{\mu}\right)
 =\mu\left(\frac{ln'}{\gamma}+\frac{l'n}{\gamma}\right)
 =\mu\Lambda\,.
 \end{equation}
 It seems that this determines a factor system $m_m^{(\Lambda)}$. However,
it is impossible to exclude a priori the case when
$\gcd(\Lambda,m)=\ell>1$. It is evident that the summands in \req{fprm} have
no common factor, but it may happen that their sum $\Lambda$ has a common
factor with $m$. Therefore, the considered product has to be decomposed into
irreducible representations with a factor system $m_M^{(\lambda)}$, where
$\lambda=\Lambda/\ell$ and $M=m/\ell$. The scalar product of appropriate
characters gives us
 \begin{eqnarray}
 f(\langle M,\lambda;\vk\rangle,
 \langle n,l;\vq\rangle\otimes\langle n',l';\vq'\rangle)
 &=&\frac{mM\gamma}{N^2}\sum_{x_1,x_2=0}^{\mu-1}
 \omega_\mu^{-(q_1+q'_1-k_1)x_1-(q_2+q'_2-k_2)x_2}
 \nonumber\\
 &=&\frac{\gamma}{\ell}\de{k_1,q_1+q'_1}\de{k_2,q_2+q'_2}\,;
 \label{freq}
 \end{eqnarray}
 there are $\ell^2$ such representations with $k_i=q_i+q'_i \bmod \mu$. 

The most interesting is the case when $n=n'$ and $l=l'$, since $n$ and $l$
are determined by the magnetic field magnitude (and the crystal period $N$).
Therefore, representations $\langle n,l;\vq\rangle$ and $\langle
n,l;\vq'\rangle$ act in two $n$-dimensional eigenspaces of one-electron
states and their product should correspond to two-electron space of states.
In this case one obtains that the resultant representation is
$n^2$-dimensional and $\gamma=m=n$, $\mu=\nu$. From \req{chiD} the character
is equal to 
\begin{equation}\label{chi2}
 \chi^D[n_1,n_2]
 =\de{n_1,x_1n}\de{n_2,x_2n} n^2
 \omega_{\nu}^{-(q_1+q'_1)x_1-(q_2+q'_2)x_2}
 \end{equation}
 with $0\le q_i,q'_i,x_i<\nu$. The factor system is given by \req{fprod}:
 \begin{equation}\label{fsq}
 m([n_1,n_2],[n'_1,n'_2])=\omega_n^{2ln_2n'_1}
 =\omega_n^{\Lambda n_2n'_1}\,,
 \end{equation}
 where, see \req{fprm}, $\Lambda=2l$. At this moment the cases of odd and 
even $n$ have to be considered separately. In the first
case $\ell=\gcd(n,2l)=1$ and the obtained representations decomposes into
$n$ copies of the representation $\langle n,2l;\vk\rangle$ with
$k_i=(q_i+q'_i)\bmod \nu$. In the second case, however, $\ell=2$ and $M=n/2$
so the considered product decomposes into representations $\langle
n/2,l;\vk\rangle$: there are four representations with $k_i-q_i-q'_i=0,\nu$
and each of them appears $n/2$ times. In a similar way, the coupling of $d$
representations $\langle n,1;\vq^{(j)}\rangle$, $j=1,2,\dots,d$ with $n=dM$
changes the magnetic period from $n$ to $M$, however it has not been caused
by modification of the magnetic field but by multiplication of the charge by
$d$ [see \req{cond}]. 

Clebsch-Gordan coefficients for the considered product are strongly
ambiguous since the frequencies of irreducible representations can be very
large so they are not discussed here. 

\section{Example}

Let us consider $N=12$ and two representations: $\langle 3,1;[1,0]\rangle$
and $\langle 6,1;[1,0]\rangle$. The corresponding co-divisors of
$n=3$ and $n'=6$ are $\nu=4$ and $\nu'=2$, respectively, so the
greatest common divisor $\gamma=3$ and the least common multiplicity
is $m=6$. Hence, a co-divisor $\mu=N/6=2$. Now we have to calculate
$L$ from Eq.~\req{fprod} and to write as a multiplicity of $\mu$, see
\req{fprm}. It is easy to obtain  that
 $$
 L=4+2=6=2\cdot3\;\Rightarrow\;\Lambda=3\,.
 $$
 So, $\ell=\gcd(\Lambda,m)=3$ and the considered product decomposes
into nine irreps. To determine them one has to find
 $$
 M={m\over\ell}={6\over 3}=2\qquad \mbox{and}
 \qquad \lambda={\Lambda\over\ell}={3\over3}=1.
 $$
  Therefore, a factor system of obtained irreducible representations
should be denoted as $m_2^{(1)}$ instead of $m_6^{(3)}$ and these
representations are two-dimensional. Due to the condition
$k_i=q_i+q'_i\bmod{\mu}$ one obtains that $k_i$, $i=1,2$, should be
even. Since $N/M=6$ then $k_i=0,1,2,3,4,5$ and, eventually, the
following decomposition can be written
 $$
 \langle 3,1;[1,0]\rangle\otimes
 \langle 6,1;[1,0]\rangle=
 \bigoplus_{k_1=0,2,4}\bigoplus_{k_2=0,2,4}
 \langle 2,1;[k_1,k_2]\rangle\,.
 $$

\subsection{Remarks}

To obtain the {\it magnetic periodicity} for $n=3$, $l=1$ and $n'=6$,
$l'=1$ one has to assume [see Eq.~\req{cond}] that
 \begin{equation}\label{hfirst}
 h=\frac{2\pi}{\Omega}{l\over n}a_3={1\over3}\frac{2\pi}{\Omega}a_3
 \end{equation}
 and
 \begin{equation}\label{hprime}
 h'=\frac{2\pi}{\Omega}{{l'}\over{n'}}a_3
 ={1\over6}\frac{2\pi}{\Omega}a_3\,.
 \end{equation}
 It seems that there are two different magnitudes of the magnetic
field since $h\neq h'$. But putting
 \begin{equation}\label{Hunique}
 h=q\frac{e}{\hbar c}H\qquad \mbox{and}\qquad
 h'=q'\frac{e}{\hbar c}H\qquad
 \end{equation}
 with
 \begin{equation}\label{twoq}
 q=2q'
 \end{equation}
 the same value of $H$ is obtained
 \begin{eqnarray*}
 H&=&h'\frac{\hbar c}{e}{1\over{q'}}
 ={1\over6}\frac{2\pi}{\Omega}a_3\frac{\hbar c}{e}{1\over{q'}}
 ={1\over3}\frac{2\pi}{\Omega}a_3\frac{\hbar c}{e}{1\over{2q'}}
 =h\frac{\hbar c}{e}{1\over q}=H
 \end{eqnarray*}
 The resultant representations correspond to $M=2$, $\lambda=1$ so
 $$
 H=\underbrace{\frac{2\pi}{\Omega}\frac{\lambda}{M}a_3}_{h''}
 \frac{\hbar c}{e}{1\over{q''}}
 ={1\over2}\frac{2\pi}{\Omega}a_3\frac{\hbar c}{e}{1\over{q''}}\,.
 $$
 Since in all three cases $H$ is the same and the charges
$q,q',q''$ have to be integers then
 $$
 {1\over{2q''}}={1\over{6q'}}={1\over{3q}}\,.
 $$
 Substituting $q=2q'$ one obtains $2q''=6q'$ and
 $$
 q''=3q'=q+q'\,,
 $$
 so multiplication of representations corresponds to addition of
charges.

\end{document}